\documentclass[aps,twocolumn,showpacs,preprintnumbers,prb]{revtex4}

\usepackage{graphicx}
\usepackage{dcolumn}
\usepackage{bm}

\begin{document}


\title{Theory of acoustic surface plasmons}
\author{J. M. Pitarke,$^{1,2}$ V. U. Nazarov,$^{3}$ V. M. Silkin,$^{2}$ E.
V. Chulkov,$^{2,4}$ E. Zaremba,$^5$ and P. M. Echenique$^{2,4}$}
\affiliation{
$^1$Materia Kondentsatuaren Fisika Saila, Zientzi Fakultatea,
Euskal Herriko Unibertsitatea,\\
644 Posta kutxatila, E-48080 Bilbo, Basque Country, Spain\\
$^2$Donostia International Physics Center (DIPC) and Centro Mixto
CSIC-UPV/EHU,\\ 
Manuel de Lardizabal Pasealekua, E-20018 Donostia, Basque Country, Spain\\
$^3$Department of Physics and Institute for Condensed Matter Theory,\\
Chonnam National University, Gwangju 500-757, Korea\\
$^4$Materialen Fisika Saila, Kimika Fakultatea, Euskal Herriko
Unibertsitatea,\\ 
1072 Posta kutxatila, E-20080 Donostia, Basque Country, Spain\\
$^5$Department of Physics, Queen's University, Kingston, Ontario, Canada K7L
3N6}

\date{June 2, 2004}

\begin{abstract}
Recently, a novel low-energy collective excitation has been predicted to
exist at metal surfaces where a quasi two-dimensional (2D) surface-state band
coexists with the underlying three-dimensional (3D) continuum. Here we present
a model in which the screening of a semiinfinite 3D metal is incorporated into
the description of electronic excitations in a 2D electron gas through the
introduction of an effective 2D dielectric function. Our self-consistent
calculations of the dynamical response of the 3D substrate indicate that an
acoustic surface plasmon exists for all possible locations of the 2D sheet
relative to the metal surface. This low-energy excitation, which exhibits
linear dispersion at low wave vectors, is dictated by the nonlocality of the
3D dynamical response providing incomplete screening of the 2D
electron-density oscillations. 
\end{abstract}


\maketitle
\section{introduction}

Since the early suggestion of Pines\cite{pines1} that low-energy plasmons with
sound-like long-wavelength dispersion could be realized in the collective
motion of a system of two types of electronic carriers, these modes have
spurred over the years a remarkable interest and research activity.\cite{tosi} 

The possibility of having a longitudinal acoustic mode in a
metal-insulator-semiconductor (MIS) structure was anticipated by
Chaplik.\cite{chaplik} Chaplik considered a simplified model in which a
two-dimensional (2D) electron gas is separated from a semiinfinite
three-dimensional (3D) metal. He found that the screening of valence electrons
in the metal changes the 2D plasmon energy from its characteristic square-root
wave-vector dependence to a linear dispersion, which was also
discussed by Gumhalter\cite{gum} in his study of transient interactions of
surface-state electron-hole (e-h) pairs at surfaces.

Nevertheless, acoustic plasmons were only expected to exist for {\it spatially
separated} plasmas, as pointed out by Das Sarma and Madhukar.\cite{sarma} The
experimental realization of two-dimensionally confined and spatially separated
multicomponent structures, such as quantum wells and heterostructures,
provided suitable solid-state systems for the observation of acoustic
plasmons.\cite{olego} Acoustic plasma oscillations were then proposed as
possible candidates to mediate the attractive interaction leading to the
formation of Cooper pairs in high-$T_c$ superconductors.\cite{ruvalds,bill}

Recently, Silkin {\it et al.}\cite{silkin} have shown that metal surfaces where
a partially occupied quasi-2D surface-state band {\it coexists} in the same
region of space with the underlying 3D continuum support a well-defined
acoustic surface plasmon, which could not be explained within the original
{\it local} model of Chaplik.\cite{chaplik} This low-energy collective
excitation exhibits linear dispersion at low wave vectors, and might therefore
affect e-h and phonon dynamics near the Fermi level.\cite{note00}

In this paper, we present a model in which the screening of a semiinfinite 3D
metal is incorporated into the description of electronic excitations in a 2D
electron gas through the introduction of an effective 2D dielectric function.
We find that the dynamical screening of valence electrons in the metal
changes the 2D plasmon energy from its characteristic square-root behaviour to
a linear dispersion, not only in the case of a 2D sheet spatially separated
from the semiinfinite metal, as anticipated by Chaplik,\cite{chaplik} but also
when the 2D sheet coexists in the same region of space with the underlying
metal, as occurs in the real situation of surface states at a metal surface.
Furthermore, our results indicate that it is the nonlocality of the
3D dynamical response which allows the formation of 2D electron-density
acoustic oscillations at metal surfaces, since these oscillations would
otherwise be completely screened by the surrounding 3D substrate.

Unless stated otherwise, atomic units are used throughout, i.e.,
$e^2=\hbar=m_e=1$.

\section{theory}

A variety of metal surfaces, such as Be(0001) and the (111) surfaces of the
noble metals Cu, Ag, and Au, are known to support a partially occupied band of
Shockley surface states with energies near the Fermi level.\cite{ingles} Since
the wavefunction of these states is strongly localized near the surface and
decays exponentially into the solid, they can be considered to form a 2D
electron gas.

In order to describe the electronic excitations occurring within a
surface-state band that is coupled with the underlying continuum of valence
electrons in the metal, we consider a model in which surface-state electrons
comprise a 2D electron gas at $z=z_d$ ($z$ denotes the coordinate normal to
the surface), while all other states of the metal comprise a 3D substrate
consisting of a fixed uniform positive background (jellium) of density
\begin{equation}
n_+(z)=\cases{\bar n,&$z\leq 0$\cr\cr
0,&elsewere,}
\end{equation}
plus a neutralizing inhomogeneous cloud of interacting electrons. The
positive-background charge density $\bar n$ is often expressed in terms of the
3D Wigner radius $r_s^{3D}=(3/4\pi\bar n)^{1/3}/a_0$, $a_0=0.529\AA$ being the
Bohr radius.

We consider the response of the interacting 2D and 3D electronic subsystems to
an external potential $\phi^{ext}({\bf r},\omega)$. According to
time-dependent perturbation theory, keeping only terms of first order in the
external perturbation, and Fourier transforming in two directions, the
electron densities induced in the 2D and 3D subsystems are found to be
\begin{eqnarray}\label{one}
&&\delta n_{2D}(z;q,\omega)=\delta(z-z_d)\,\chi_{2D}(q,\omega)
\cr\cr
&\times&\left[\phi^{ext}(z;q,\omega)+\int dz'v(z,z';q)\delta
n_{3D}(z';q,\omega)\right]
\end{eqnarray}
and
\begin{eqnarray}\label{two}
&&\delta n_{3D}(z;q,\omega)=\int
dz'\,\chi_{3D}(z,z';q,\omega)\cr\cr
&\times&\left[\phi^{ext}(z';q,\omega)+\int dz''v(z',z'';q)\delta
n_{2D}(z'';q,\omega)\right].
\end{eqnarray}
Here, $q$ is the magnitude of the 2D wave vector parallel to the surface,
$\chi_{2D}(q,\omega)$ and $\chi_{3D}(z,z';q,\omega)$ are 2D
and 3D {\rm interacting} density response functions, respectively,
$\phi^{ext}(z;q,\omega)$ is the 2D Fourier transform of the external potential
$\phi^{ext}({\bf r},\omega)$, and $v(z,z';q)$ is the 2D Fourier transform of
the bare Coulomb interaction:
\begin{equation}\label{bare}
v(z,z';q)=v_q\,{\rm e}^{-q\,|z-z'|},
\end{equation}
with $v_q=2\pi/q$. 

Combining Eqs.~(\ref{one}) and (\ref{two}), we find
\begin{equation}\label{eff1}
\delta n_{2D}(z;q,\omega)=
\delta(z-z_d)\,\chi_{eff}(q,\omega)\,\tilde\phi(z;q,\omega),
\end{equation}
where
\begin{equation}\label{chieff}
\chi_{eff}(q,\omega)={\chi_{2D}(q,\omega)\over
1-\chi_{2D}(q,\omega)\left[W(z_d,z_d;q,\omega)-v_q\right]},
\end{equation}
$W(z_d,z_d;q,\omega)$ being the so-called screened interaction
\begin{eqnarray}\label{w}
&&W(z,z';q,\omega)=v(z,z';q)+\int dz_1\int dz_2\cr\cr
&&\times\,v(z,z_1;q)\,\chi_{3D}(z_1,z_2;q,\omega)\,v(z_2,z';q),
\end{eqnarray}
and $\tilde\phi(z;q,\omega)$ being the 2D Fourier transform of the total
potential at $z$ in the absence of the 2D sheet:
\begin{eqnarray}\label{phitilde}
\tilde\phi(z;q,\omega)&=&\int dz''\left[\delta(z-z'')+\int
dz'\,v(z,z';q)\right.\cr\cr
&\times&\left.\chi_{3D}(z',z'';q,\omega)\right]\phi^{ext}(z'';q,\omega).
\end{eqnarray}
Eq.~(\ref{eff1}) suggests that the screening of the 3D subsystem can be
incorporated into the description of the electron-density response at the 2D
electron gas through the introduction of the effective density-response
function of Eq.~(\ref{chieff}), whose poles should correspond to 2D
electron-density oscillations.

Alternatively, we can focus on the 2D Fourier transform of the total potential
at $z$ in the presence of both 2D and 3D subsystems:
\begin{eqnarray}\label{six}
\phi(z;q,\omega)&=&\phi^{ext}(z;q,\omega)+\int dz'\,v(z,z';q)\cr\cr
&\times&\left[\delta n_{2D}(z';q,\omega)+\delta n_{3D}(z';q,\omega)\right],
\end{eqnarray}  
which with the aid of Eqs.~(\ref{two}) and (\ref{phitilde}) can also be
expressed in the following way: 
\begin{equation}
\phi(z;q,\omega)=\tilde\phi(z;q,\omega)+\int dz'W(z,z';q,\omega)\,\delta
n_{2D}(z';q,\omega).
\end{equation}
Now we choose $z=z_d$, and using Eq.~(\ref{eff1}) we write
\begin{equation}\label{seven}
\phi(z_d;q,\omega)=\left[1+W(z_d,z_d;q,\omega)\,\chi_{eff}(q,\omega)\right]\,
\tilde\phi(z_d;q,\omega),
\end{equation}
which allows to introduce the effective inverse 2D dielectric function
\begin{equation}\label{eff}
\epsilon_{eff}^{-1}(q,\omega)=1+W(z_d,z_d;q,\omega)\,\chi_{eff}(q,\omega).
\end{equation}

Since our aim is to investigate the occurrence of long-wavelength ($q\to 0$)
collective excitations, we can rely on the random-phase approximation
(RPA),\cite{pines2} which is exact in the $q\to 0$ limit. In this
approximation, the 2D and 3D interacting density-response functions are
obtained as follows
\begin{equation}\label{rpa2d}
\chi_{2D}(q,\omega)={\chi_{2D}^0(q,\omega)\over 1-\chi_{2D}^0(q,\omega)\,v_q}
\end{equation}
and
\begin{eqnarray}\label{rpa}
&&\chi_{3D}(z,z';q,\omega)=\chi_{3D}^0(z,z';q,\omega)+\int dz_1\int dz_2\cr\cr
&&\times\,\chi_{3D}^0(z,z_1;q,\omega)\,v(z_1,z_2;q)\,\chi_{3D}(z_2,z';q,\omega),
\end{eqnarray}
where $\chi_{2D}^0(q,\omega)$ and $\chi_{3D}^0(z,z';q,\omega)$ represent their
noninteracting counterparts. An explicit expression for the 2D noninteracting
density-response function $\chi_{2D}^0(q,\omega)$ was reported by
Stern.\cite{stern} In order to derive explicit expressions for the 3D
noninteracting density-response function $\chi_{3D}^0(z,z';q,\omega)$ one needs
to rely on simple models, such as the hydrodynamic or infinite-barrier model,
but accurate numerical calculations have been carried
out\cite{eguiluz,liebsch} from the knowledge of the eigenfunctions and
eigenvalues of the Kohn-Sham hamiltonian of density-functional theory
(DFT).\cite{dft}

Combining Eqs.~(\ref{chieff}), (\ref{eff}), and (\ref{rpa2d}), one finds the
RPA effective 2D dielectric function
\begin{equation}\label{eff2}
\epsilon_{eff}(q,\omega)=1-W(z_d,z_d;q,\omega)\,\chi_{2D}^0(q,\omega).
\end{equation}
The longitudinal modes of the 2D subsystem, or plasmons, are solutions of
\begin{equation}\label{zero}
\epsilon_{eff}(q,\omega)=0.
\end{equation}

In the absence of the 3D subsystem, the 3D screened interaction
$W(z,z';q,\omega)$ reduces to the bare Coulomb interaction $v(z,z';q)$, and the
solution of Eq.~(\ref{zero}) leads at long
wavelengths to the well-known square-root wave-vector dependence of
the 2D plasmon energy\cite{stern}
\begin{equation}\label{2d}
\omega_{2D}={q_F\over\sqrt{m}}\,\sqrt{q},
\end{equation}
$q_F$ and $m$ being the 2D Fermi momentum and 2D effective mass, respectively.
The 2D Fermi velocity is simply $v_F=q_F/m$.

In the presence of the 3D subsystem, the long-wavelength limit of the effective
2D dielectric function of Eq.~(\ref{eff2}) is found to have two zeros. One
zero corresponds to a high-frequency ($\omega>>v_Fq$) oscillation in which 2D
and 3D electrons oscillate in phase with one another. The other mode
corresponds to a low-frequency acoustic oscillation in which both 2D and 3D
electrons oscillate out of phase. 

At high frequencies, where $\omega>>v_Fq$, the long-wavelength limit of
the 2D density-response function $\chi_{2D}^0(q,\omega)$ is known to be
\begin{equation}\label{high1}
\lim_{q\to 0}\chi_{2D}^0(q,\omega>>v_Fq)={1\over v_q}
\,{\omega_{2D}^2\over\omega^2}.
\end{equation}
On the other hand, when the 2D sheet is located either far inside or far
outside the metal surface, the long-wavelength limit of the 3D screened
interaction $W(z_d,z_d;q,\omega)$ takes the form
\begin{equation}\label{high2}
\lim_{q\to
0}W(z_d,z_d;q,\omega>>v_Fq)=v_q\,{\omega^2\over\omega^2-\omega_{p,s}^2},
\end{equation}
where $\omega_{p,s}$ represents either the bulk-plasmon frequency
$\omega_p=\sqrt{4\,\pi\,\bar n}$ or the conventional surface-plasmon energy
$\omega_s=\omega_p/\sqrt 2$,\cite{ritchie} depending on whether the 2D sheet
is located inside or outside the solid. Introduction of Eqs.~(\ref{high1}) and
(\ref{high2}) into Eqs.~(\ref{eff2}) and (\ref{zero}) yields a high-frequency
mode at
\begin{equation}
\omega^2=\omega_{p,s}^2+\omega_{2D}^2.
\end{equation}

At low frequencies, we seek for an acoustic 2D plasmon energy that in the
long-wavelength limit takes the form
\begin{equation}\label{ac}
\omega=\alpha\,v_F\,q.
\end{equation}
A careful analysis of the 2D density-response function $\chi_{2D}^0(q,\omega)$
and the 3D screened interaction $W(z_d,z_d;q,\omega)$ shows that at
$\omega=\alpha\,v_F\,q$ the long-wavelength limits of these quantities take
the form
\begin{equation}\label{limit1}
\lim_{q\to 0}\chi_{2D}^0(q,\alpha
v_Fq)={1\over\pi}\left[{\alpha\over\sqrt{\alpha^2-1}}-1\right]
\end{equation}
and
\begin{equation}\label{limit2}
\lim_{q\to 0}W(z_d,z_d;q,\alpha v_Fq)=I(z_d).
\end{equation}
An inspection of Eqs.~(\ref{eff2}), (\ref{limit1}), and (\ref{limit2})
indicates that for a low-energy acoustic oscillation to occur the quantity
$I(z_d)$ must be different from zero. In that case, the long-wavelength limit
of the effective 2D dielectric function of Eq.~(\ref{eff2}) has indeed a zero
corresponding to a low-frequency oscillation of energy given by Eq.~(\ref{ac})
with
\begin{equation} \label{alpha2}
\alpha=\sqrt{1+{\left[I(z_d)\right]^2\over\pi\left[\pi+2\,I(z_d)\right]}}.
\end{equation}

In the following, we investigate the impact of the 3D screening on the actual
wave-vector dependence of the low-energy 2D collective excitation. We first
consider the two limiting cases in which the 2D sheet is located far inside
and far outside the metal surface, and we then carry out self-consistent
calculations of the 3D screened interaction $W(z,z';q,\omega)$, which will
allow us to obtain plasmon dispersions for arbitrary locations of the 2D
sheet.

\subsection{2D sheet far inside the metal surface}

In the case of a 2D sheet that is located far inside the metal surface, the 3D
subsystem can safely be assumed to exhibit translational invariance in all
directions, i.e., the screened interaction $W(z_d,z_d;q,\omega)$ entering
Eq.~(\ref{eff2}) can be easily obtained from the knowledge of the interacting
density-response function $\chi_{3D}(k,\omega)$  of a uniform 3D electron gas,
as follows
\begin{equation}\label{bulk}
W(z_d,z_d;q,\omega)=2\int{dq_z\over k^2}\,\epsilon_{3D}^{-1}(k,\omega),
\end{equation}
where $k=\sqrt{q^2+q_z^2}$ is the magnitude of a 3D wave vector and
$\epsilon_{3D}^{-1}(k,\omega)$ is the inverse dielectric function of a uniform
3D electron gas:
\begin{equation}
\epsilon_{3D}^{-1}(k,\omega)=1+{4\pi\over k^2}\,\chi_{3D}(k,\omega).
\end{equation}
In the RPA,
\begin{equation}
\epsilon_{3D}(k,\omega)=1-{4\pi\over k^2}\,\chi_{3D}^0(k,\omega),
\end{equation}
$\chi_{3D}^0(k,\omega)$ being the noninteracting density-response function
first obtained by Lindhard.\cite{lindhard} 

\subsubsection{Local 3D response}

If one characterizes the 3D uniform electron gas by a local
dielectric function $\epsilon_{3D}(\omega)$, then Eq.~(\ref{bulk}) yields 
\begin{equation}\label{local}
W^{local}(z_d,z_d;q,\omega)=v_q\,\epsilon_{3D}^{-1}(\omega).
\end{equation}
In a 3D gas of free electrons, $\epsilon_{3D}(\omega)$ takes the Drude form
\begin{equation}\label{drude}
\epsilon_{3D}(\omega)=1-{\omega_p^2\over\omega^2},
\end{equation}
which yields
\begin{equation}\label{limit3}
\lim_{q\to 0}W^{local}(z_d,z_d;q,\alpha v_Fq)=0.
\end{equation}
This means that in a local picture of the 3D response the characteristic
collective oscillations of the 2D electron gas would be completely screened by
the sorrounding 3D substrate and no low-energy acoustic mode would
exist.\cite{note01} 

\subsubsection{Hydrodynamic 3D response}

Dispersion effects of the 3D subsystem can be incorporated approximately in a
hydrodynamic model. In this approximation, the dielectric
function of a 3D uniform electron gas is found to be\cite{lindhard}
\begin{equation}\label{hydro}
\epsilon_{3D}(k,\omega)=1-{\omega_p^2\over\omega^2-\beta^2\,k^2},
\end{equation}
where $\beta=\sqrt{1/3}\,k_F$ represents the speed of propagation of
hydrodynamic disturbances in the electron system,\cite{note0} and $k_F$ is the
3D Fermi momentum.

Introducing Eq.~(\ref{hydro}) into Eq.~(\ref{bulk}), one finds
\begin{equation}\label{limit4}
\lim_{q\to 0}W(z_d,z_d;q,\alpha v_Fq)=2\pi\beta/\omega_p,
\end{equation}
which yields the following simple expression for the acoustic coefficient of
Eq.~(\ref{alpha2}):
\begin{equation}\label{alpha}
\alpha=\sqrt{1+{4\beta^2/\omega_p^2\over1+4\beta/\omega_p}}.
\end{equation}

\subsubsection{Full 3D response}

We have carried out numerical calculations of the RPA effective dielectric
function of Eq.~(\ref{eff2}), by using the full $\chi_{2D}^0(q,\omega)$ and
$\chi_{3D}^0(k,\omega)$ density-response functions, and choosing the
electron-density parameters $r_s^{2D}=3.14$ and $r_s^{3D}=1.87$ corresponding
to the (0001) surface of Be.\cite{note1}

\begin{figure}
\includegraphics[width=0.45\textwidth,height=0.3375\textwidth]{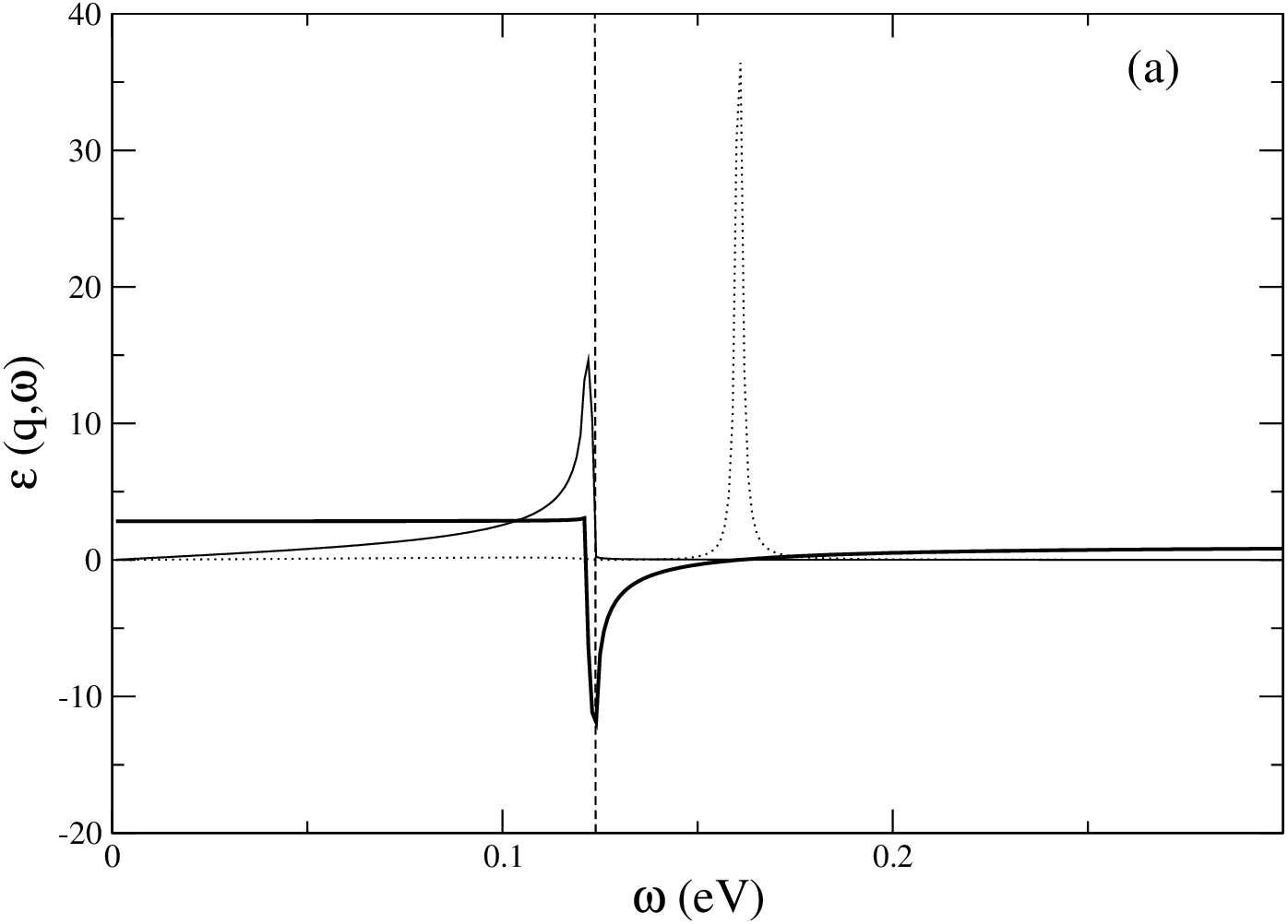}\\
\vspace*{1.2cm}
\includegraphics[width=0.45\textwidth,height=0.3375\textwidth]{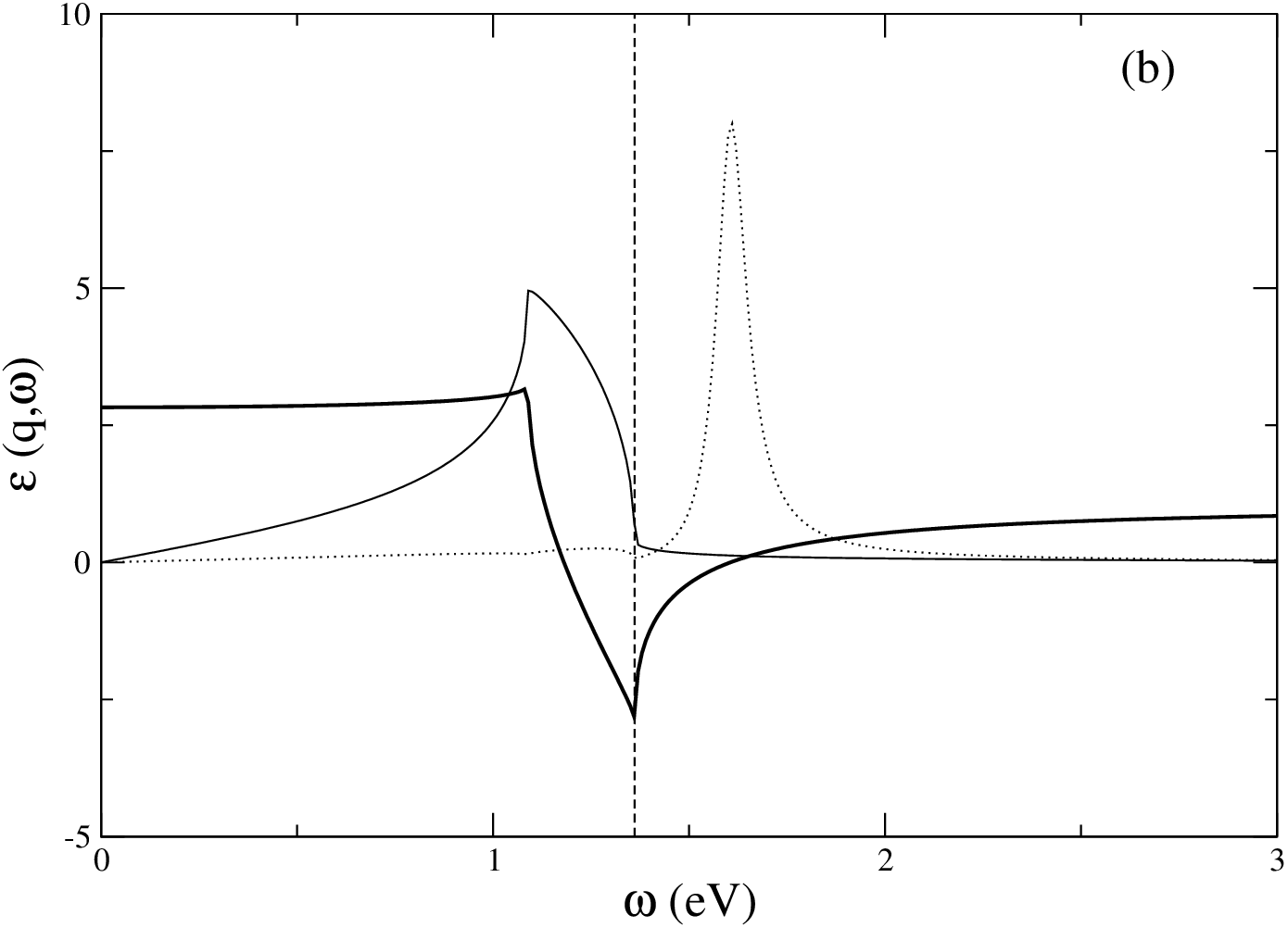}\\
\caption{Effective dielectric function of a 2D sheet that is located far inside
the metal surface, as obtained from Eq.~(\ref{eff2}) with (a) $q=0.01$ and (b)
$q=0.1$. The real and the imaginary parts of $\epsilon_{eff}(q,\omega)$ are
represented by thick and thin solid lines, respectively. The dotted line
represents the effective 2D energy-loss function ${\rm
Im}\left[-\epsilon_{eff}^{-1}(q,\omega)\right]$.
The vertical dashed line represents the upper edge $\omega_u=v_Fq+q^2/(2m)$ of the
2D e-h pair continuum, where 2D e-h pairs can be excited. 2D and 3D electron
densities have been taken to be those corresponding to the Wigner radii
$r_s^{2D}=3.14$ and $r_s^{3D}=1.87$, respectively. The 2D effective mass has
been taken to be $m=1$.\label{fig1}}
\end{figure}

The results we have obtained with $q=0.01a_0^{-1}$ and $q=0.1a_0^{-1}$ are
displayed in Figs.~\ref{fig1}a and \ref{fig1}b, respectively. We observe that
at energies below the upper edge $\omega_u=v_Fq+q^2/(2m)$ (vertical dashed line)
of the 2D e-h pair
continuum (where 2D e-h pairs can be excited) the real
part of the effective dielectric function is nearly constant and the imaginary
part is large, as would occur in the absence of the 3D susbtrate. At energies
above $\omega_u$, momentum and energy conservation prevents 2D e-h pairs from
being produced, and ${\rm
Im}\,\epsilon_{eff}(q,\omega)$ is very small.

Collective excitations are related to a zero of ${\rm
Re}\,\epsilon_{eff}(q,\omega)$ in a region where ${\rm
Im}\,\epsilon_{eff}(q,\omega)$ is small and lead, therefore, to a maximum in
the energy-loss function ${\rm Im}[-\epsilon_{eff}^{-1}(q,\omega)]$.\cite{loss}
In the absence of the 3D substrate, a 2D plasmon would occur at
$\omega_{2D}=1.22\,{\rm eV}$ for $q=0.01a_0^{-1}$ and $\omega_{2D}=3.99\,{\rm
eV}$ for $q=0.1a_0^{-1}$. However, Fig.~\ref{fig1} shows that in the presence
of the 3D substrate a well-defined low-energy {\it acoustic} plasmon occurs,
the sound velocity being just over the 2D Fermi velocity $v_F$. The small
width of the plasmon peak is entirely due to plasmon decay into e-h pairs of
the 3D substrate.

\begin{figure}
\includegraphics[width=0.45\textwidth,height=0.3375\textwidth]{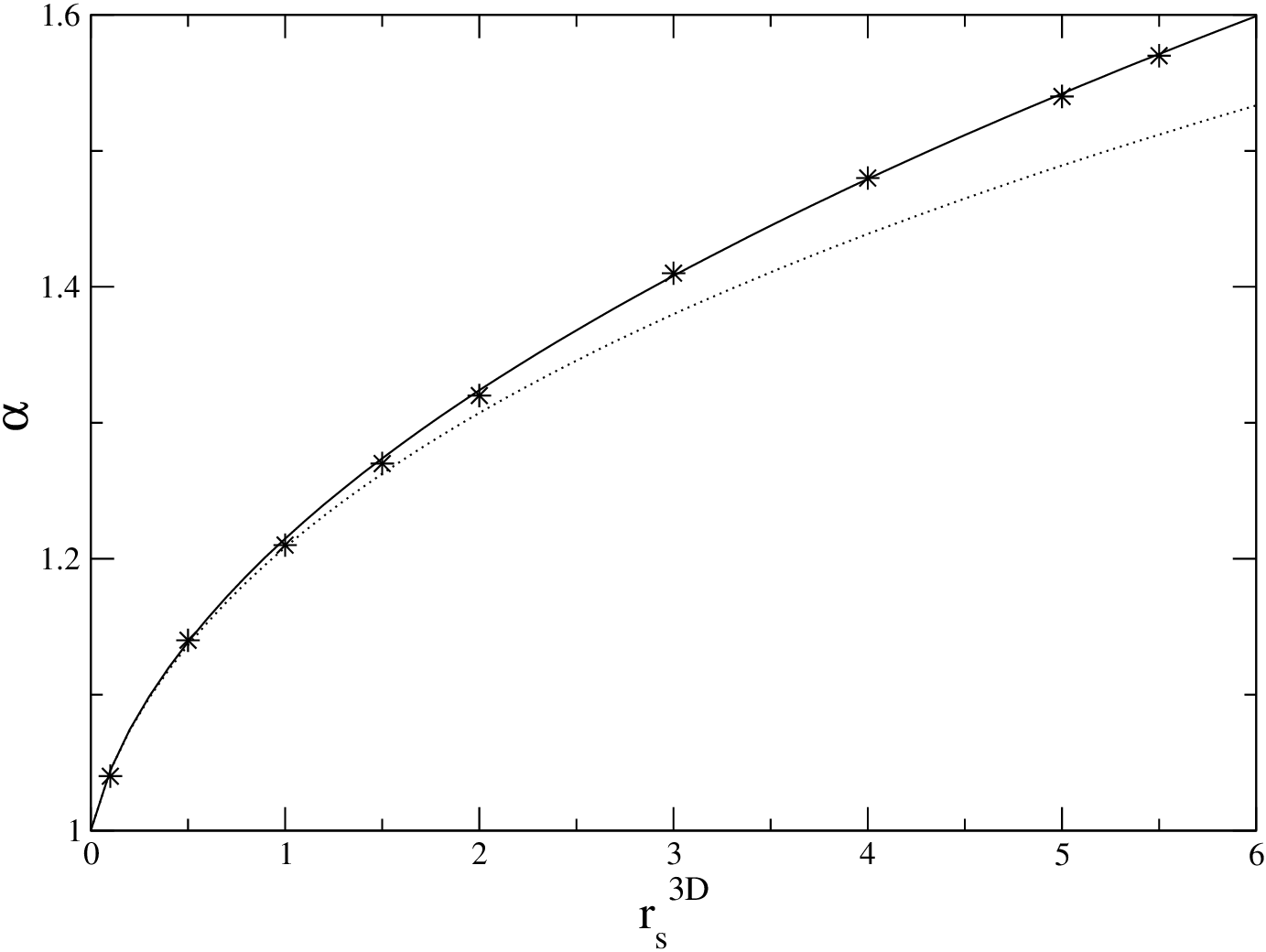}
\caption{Stars represent the $\alpha$ coefficient of the acoustic-plasmon
energy $\omega=\alpha\,v_F\,q$ versus the 3D Wigner radius, as obtained from
Eq.~(\ref{zero}) in the long-wavelength limit. These results are found to be
insensitive to the 2D Wigner radius $r_s^{2D}$. The solid line represents the
prediction of Eq.~(\ref{alpha2}), as obtained with the full RPA value of
$I(z_d\to-\infty)$. The dotted line represents the hydrodynamic prediction of
Eq.~(\ref{alpha}).\label{fig2}}
\end{figure}

We have carried out calculations of the effective 2D dielectric function of
Eq.~(\ref{eff2}) for a variety of 2D and 3D electron densities, and we have
found that a well-defined acoustic plasmon of energy $\omega=\alpha\,v_F\,q$
is always present for 2D wave vectors up to a maximum value of $q\sim q_F$
where the acoustic-plasmon dispersion merges with $\omega_u$. The coefficient
$\alpha$ that we have obtained from the zeros in Eq.~(\ref{zero}) is
represented by stars in Fig.~\ref{fig2} versus the 3D Wigner radius
$r_s^{3D}$, together with the prediction of Eq.~(\ref{alpha2}) as obtained
with the computed RPA value of $I(z_d\to-\infty)$ (solid line) and the
hydrodynamic prediction of Eq.~(\ref{alpha}) (dotted line).
Fig.~\ref{fig2} shows that Eq.~(\ref{alpha}) is a good representation of the
linear dispersion of this low-energy plasmon, especially at the highest 3D
electron densities. Fig.~\ref{fig2} also shows that for low electron densities the hydrodynamic prediction is too small, which is due to the fact that at low densities the long-wavelength limit of the 3D screened interaction is underestimated in this approximation.

\subsection{2D sheet far outside the metal surface}

In the case of a 2D sheet that is located far outside the metal surface, where
the 3D electron density is negligible, the 3D screened interaction of
Eq.~(\ref{w}) at $z=z'=z_d$ takes the form
\begin{equation}\label{g}
W(z_d,z_d;q,\omega)=v_q\left[1-{\rm e}^{-2\,q\,z_d}\,g(q,\omega)\right],
\end{equation}
where $g(q,\omega)$ is the so-called surface-response function of the 3D
subsystem
\begin{equation}\label{g2}
g(q,\omega)=-v_q\int dz_1\int dz_2\,{\rm
e}^{q\,(z_1+z_2)}\,\chi_{3D}(z_1,z_2;q,\omega).
\end{equation}

\subsubsection{Local 3D response}

In the simplest possible model of a metal surface, one characterizes the 3D
substrate at $z\leq 0$ by a local dielectric function which jumps
discontinuously at the surface from $\epsilon_{3D}(\omega)$ inside the metal
($z\leq 0$) to zero outside ($z>0$). Witin this model,\cite{liebsch2}
\begin{equation}\label{glocal}
g^{local}(q,\omega)={\epsilon_{3D}(\omega)-1\over\epsilon_{3D}(\omega)+1},
\end{equation}
which is precisely the long-wavelength limit of the actual surface-response
function. 

At low frequencies, where $\epsilon_{3D}(\omega)$ is large [see
Eq.~(\ref{drude})] and $g^{local}(q,\omega)\to 1$, Eq.~(\ref{g}) yields
\begin{equation}\label{wout}
\lim_{q\to 0}W^{local}(z_d,z_d;q,\alpha v_Fq)=4\,\pi\,z_d.
\end{equation}
Introducing Eq.~(\ref{wout}) into Eq.~(\ref{alpha2}), one finds
\begin{equation} \label{alpha3}
\alpha=\sqrt{1+{16\,z_d^2\over 1+8\,z_d}}.
\end{equation}
For large values of the distance $z_d$ between the 2D sheet and the metal
surface, one can write
\begin{equation}\label{local2}
\alpha\approx\sqrt{2z_d},
\end{equation}
which is the result first obtained by Chaplik\cite{chaplik} by using the
Drude-like 2D density-response function of Eq.~(\ref{high1}).

\subsubsection{Nonlocal 3D response}

An inspection of Eq.~(\ref{g}) shows that the long-wavelength limit of the
screened interaction $W(z_d,z_d;q,\omega)$ is dictated not only by the local
($q=0$) surface-response function $g^{local}(q,\omega)$ but also by the
leading correction in $q$ of the actual nonlocal $g(q,\omega)$. Feibelman
showed that up to first
order in an expansion in powers of $q$, the surface-response function of a
jellium surface can be written as\cite{feibelman}
\begin{equation}\label{expansion}
g(q,\omega)={\epsilon_{3D}(\omega)-1\over\epsilon_{3D}(\omega)+1}\left[1+2qd_\perp(\omega)
{\epsilon_{3D}(\omega)\over\epsilon_{3D}(\omega)+1}\right]+O(q^2),
\end{equation}
which at low frequencies yields
\begin{equation}\label{expansion2}
g(q,\omega)\approx 1+2\,q\,d_\perp(0).
\end{equation}
The frequency-dependent $d_\perp(\omega)$ function occurring in
Eq.~(\ref{expansion}) represents the centroid of the induced 3D charge density,
which in the static limit ($\omega=0$) reduces to the image plane of an
external point charge.

Using Eq.~(\ref{expansion2}), we find the actual long-wavelength limit of
Eq.~(\ref{g}):
\begin{equation}\label{wout2}
\lim_{q\to 0}W(z_d,z_d;q,\alpha v_Fq)=4\,\pi\,\left[z_d-d_\perp(0)\right],
\end{equation}
which combined with Eq.~(\ref{alpha2}) yields
\begin{equation} \label{alpha4}
\alpha=\sqrt{1+{16\,\left[z_d-d_\perp(0)\right]^2\over
1+8\,\left[z_d-d_\perp(0)\right]}}.
\end{equation}
This shows that the acoustic-plasmon sound velocity derived from the local
model [see Eq.~(\ref{alpha3})] remains unchanged, as long as $z_d$ is replaced
by the coordinate of the 2D sheet relative to the position of the image plane.  

\subsubsection{Full 3D response}

In order to compute the full RPA surface-response function of Eq.~(\ref{g2}),
we follow the method described in Ref.~\onlinecite{eguiluz} for a jellium
slab. We first assume that the 3D electron density vanishes at a distance
$z_0$ from either jellium edge,\cite{z0} and compute the noninteracting
density-response function $\chi_{3D}^0(z,z';q,\omega)$ from the knowledge of
the self-consistent Kohn-Sham wavefunctions and energies of DFT,\cite{dft}
which we obtain in the local-density approximation
(LDA).\cite{lda} We then introduce a double-cosine
Fourier representation for both the noninteracting and the interacting
density-response functions, and find explicit expressions for the
surface-response function in terms of the Fourier coefficients of the
density-response function.\cite{aran} To ensure that our slab calculations are
a faithful representation of the actual surface-response function of a
semiinfinite 3D system, we follow the extrapolation procedure described
in Ref.~\onlinecite{pitarke}.

\begin{figure}
\includegraphics[width=0.45\textwidth,height=0.3375\textwidth]{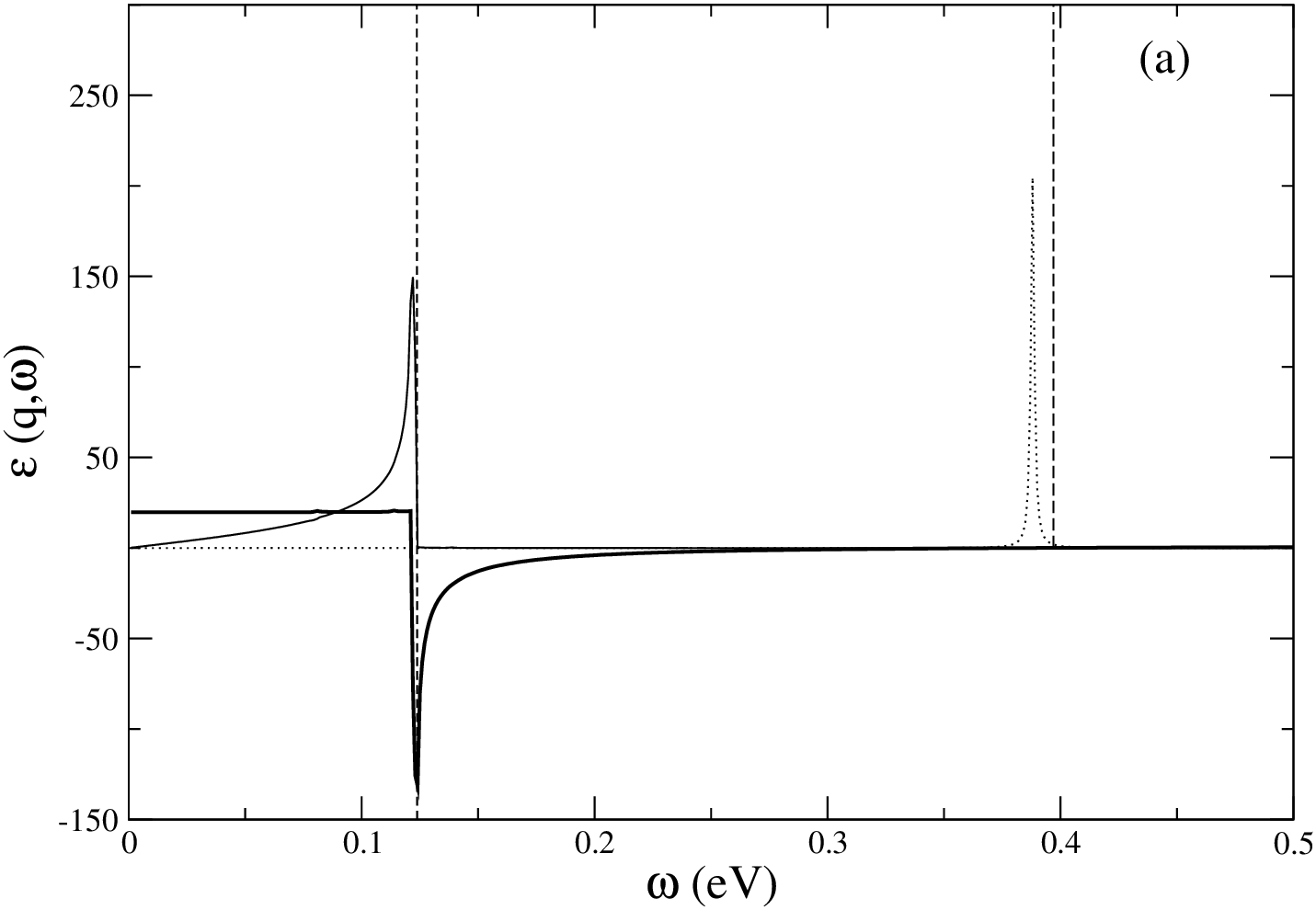}\\
\vspace*{1.2cm}
\includegraphics[width=0.45\textwidth,height=0.3375\textwidth]{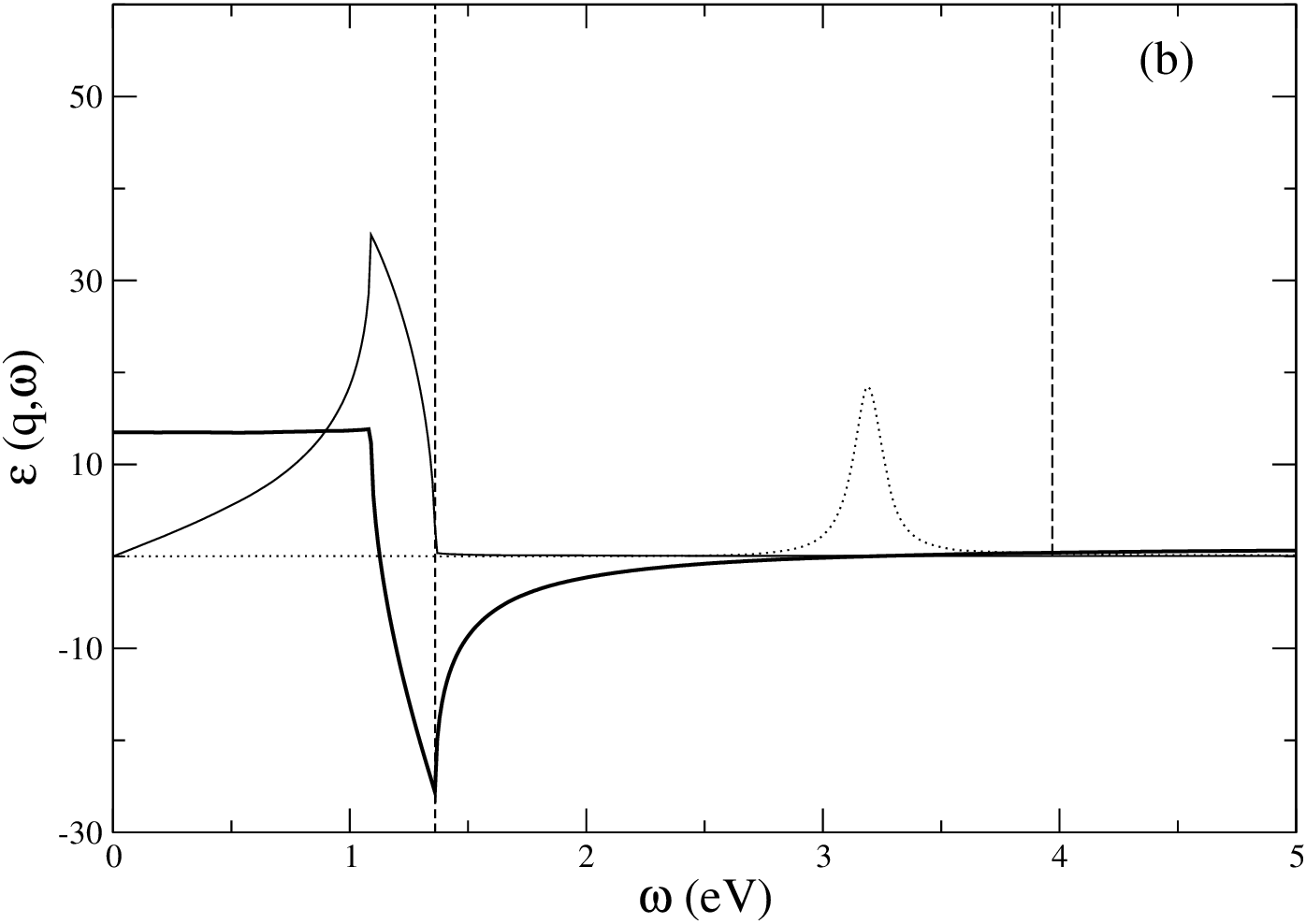}\\
\caption{As in Fig.~\ref{fig1}, but now for a 2D sheet that is located at one
3D Fermi wavelength outside the metal surface ($z_d=\lambda_F$). The
long-dashed vertical lines here represent the plasmon-energy
$\omega=\alpha\,v_F\,q$ predicted by Eq.~(\ref{alpha4}) with
$d_\perp=0.2\,\lambda_F$. For real frequencies, a 2D sheet that is located at
$z_d=\lambda_F$ exhibits a plasmon peak that at $q=0.01\,a_0^{-1}$ is
extremely sharp (as $z_d\to\infty$ the plasmon peak becomes a delta function);
hence, in the calculations presented in this figure we have replaced the
energy $\omega$ by a complex quantity $\omega+i\eta$ with (a) $\eta=0.05\,{\rm
eV}$ for $q=0.01\,a_0^{-1}$ and (b) $\eta=0$ for
$q=0.1\,a_0^{-1}$.\label{fig3}}
\end{figure}

We have carried out numerical calculations of the effective dielectric function
of Eq.~(\ref{eff2}), by using the full 2D noninteracting density-response
function, $\chi_{2D}^0(q,\omega)$, and the self-consistent RPA
surface-response function, $g(q,\omega)$, with electron-density parameters
$r_s^{2D}=3.14$ and
$r_s^{3D}=1.87$ corresponding to Be(0001).

The results we have obtained for a
2D sheet located at $z_d=\lambda_F$ are
displayed in Figs.~\ref{fig3}a (with $q=0.01a_0^{-1}$) and \ref{fig3}b (with
$q=0.1a_0^{-1}$),
$\lambda_F=2\pi/k_F$ being the 3D Fermi wavelength. Fig.~\ref{fig3} clearly
shows that in the presence of the 3D substrate a well-defined low-energy {\it
acoustic} plasmon occurs, the sound velocity being close to that predicted by
Eq.~(\ref{alpha4}) with $d_\perp(0)=0.2\,\lambda_F$ (vertical long-dashed
lines). The actual plasmon energy is smaller than predicted by
Eq.~(\ref{alpha4}), especially at the shortest wavelengths ($q=0.1a_0^{-1}$),
simply due to the bending of the plasmon dispersion as a function of $q$ (see
Fig.~\ref{fig6} below).

\subsection{2D sheet at an arbitrary location}

\subsubsection{Hydrodynamic 3D response}

An explicit expression for the screened interaction $W(z,z';q,\omega)$ of
Eq.~(\ref{w}) can be obtained in a hydrodynamic model in which the 3D electron
density is assumed to change abruptly at the surface from $\bar n$ inside the
metal to zero outside. After writing and linearizing the basic hydrodynamic
equations, i.e., the continuity and the Bernouilli equation, we find
\begin{equation}\label{lh}
\lim_{q\to 0}W(z_d,z_d;q,\alpha v_Fq)=\cases{{2\pi\beta\over\omega_p}
\left(1-{\rm e}^{2z_d\omega_p/\beta}\right)&$z_d\leq 0$\cr\cr
4\,\pi\,z_d&$z_d>0$},
\end{equation}
which combined with Eq.~(\ref{alpha2}) yields an explicit expression for the
acoustic coefficient $\alpha$. We note that in a local description of the electronic response of the solid surface ($\beta=0$) the 3D screened interaction $W(z_d,z_d;q,\alpha v_Fq)$ is zero inside the solid ($z_d\leq 0$) and $4\,\pi\,z_d$ outside ($z_d>0$). This shows that in the 2D
long-wavelength limit ($q\to 0$) the nonlocality of the 3D response is only
present inside the solid ($z_d\leq 0$), where finite values of the 3D momentum
$k$ are possible.   

Alternatively, the screened interaction $W(z,z';q,\omega)$ can be obtained
within a specular-reflection model (SRM)\cite{ritchie2} or,
equivalently, a classical infinite-barrier model (CIBM)\cite{griffin,nazarov}
of the surface, which have the virtue of describing the 3D screened
interaction in terms of the bulk dielectric function
$\epsilon_{3D}(k,\omega)$ of a 3D uniform (and infinite) electron gas (see
Appendix~\ref{ap2}). If this bulk dielectric function is chosen to be the
hydrodynamic dielectric function of Eq.~(\ref{hydro}), then these models yield
Eq.~(\ref{lh}). A more accurate
description of the bulk dielectric function $\epsilon_{3D}(k,\omega)$ yields a
result that still coincides with that of Eq.~(\ref{lh}) outside the surface
($z_d>0$), though small differences may arise at $z_d<0$.

When the 2D sheet is located far inside the metal ($z_d<<0$),
Eq.~(\ref{lh}) yields the hydrodynamic asymptotic behaviour dictated by
Eq.~(\ref{limit4}), and the SRM combined with the RPA bulk dielectric function
yields the correct RPA asymptotic behaviour. However, these hydrodynamic and
specular-reflection models, which are both based on the
assumption that the 3D electron density drops abruptly to zero at the surface,
fail to reproduce the correct asymptotic behaviour outside the surface [see
Eq.~(\ref{wout2})]. This is due to the fact that the leading correction in $q$
of the surface-response function $g(q,\omega)$ is governed by the
spill out of the electron density into the vacuum, which is not present in
these models.

\subsubsection{Full 3D response}

For an arbitrary location of the 2D sheet we need to compute the full screened
interaction $W(z_d,z_d;q,\omega)$ of Eq.~(\ref{w}). To calculate this quantity
we consider a jellium slab, as we did to obtain the surface-response function
$g(q,\omega)$, and we find explicit expressions in terms of the Fourier
coefficients of the interacting density-response function,\cite{aran} which
we  compute in the RPA [see Eq.~(\ref{rpa})] from the knowledge of the LDA
eigenvalues and eigenfunctions of the Kohn-Sham hamiltonian of DFT.

\begin{figure}
\includegraphics[width=0.45\textwidth,height=0.3375\textwidth]{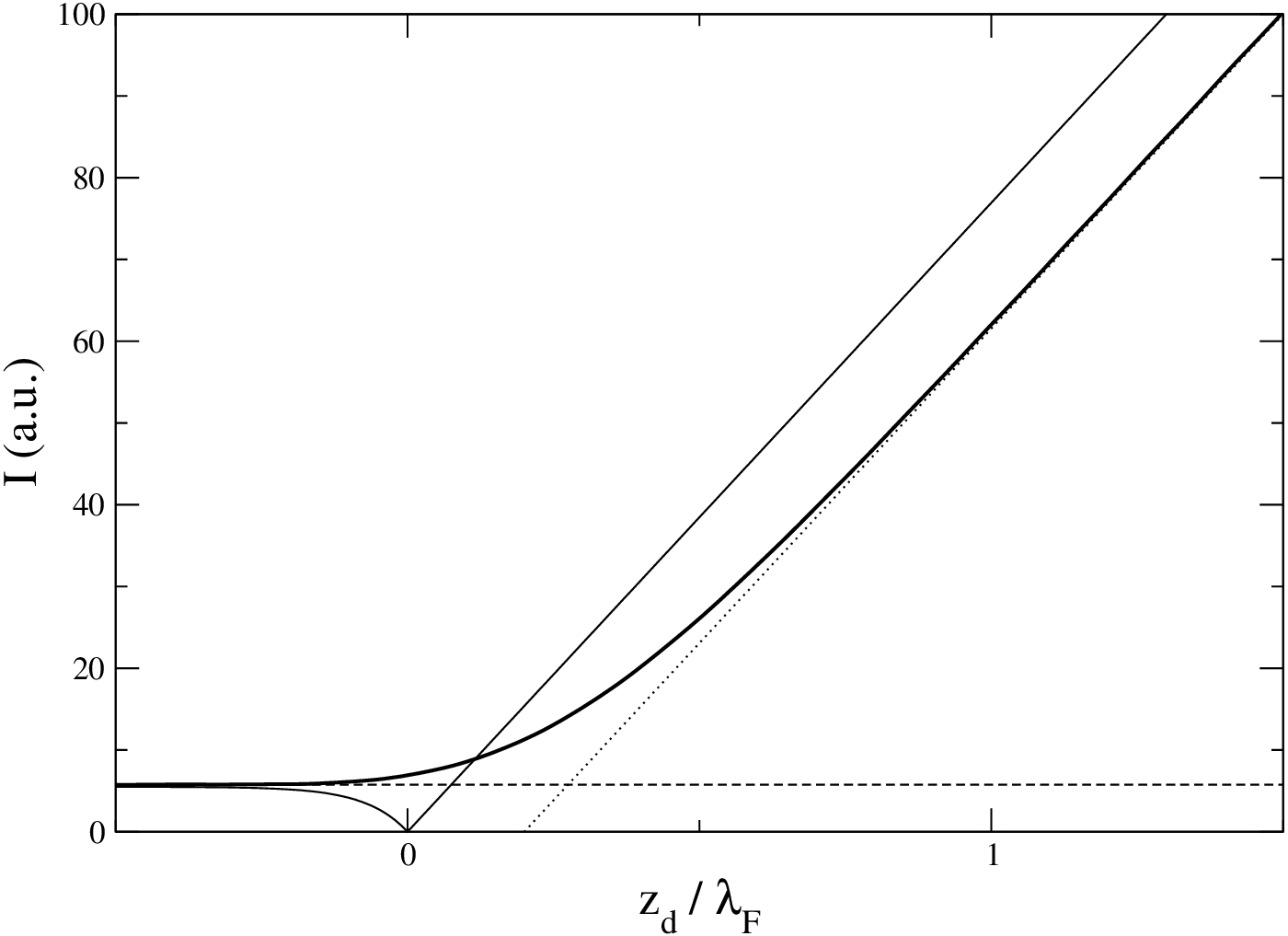}\\
\caption{Long-wavelength limit $I(z_d)$ of the screened interaction
$W(z_d,z_d;q,\alpha v_Fq)$. The thick solid line represents the full
self-consistent RPA calculation. The results obtained from Eq.~(\ref{wout2})
with $d_\perp(0)=0.2\,\lambda_F$ and from Eq.~(\ref{lh}) are represented by
dotted and thin solid lines, respectively. The horizontal dashed line
represents the result obtained from the RPA bulk screened interaction of
Eq.~(\ref{bulk}). The 3D Wigner radius has been taken to be $r_s^{3D}=1.87$.
\label{fig4new}}
\end{figure}

In Fig.~\ref{fig4new}, the long-wavelength limit $I(z_d)$ of the screened
interaction $W(z_d,z_d;q,\alpha v_Fq)$ [see Eq.~(\ref{limit2})] is displayed
versus $z_d$, as obtained with $r_s^{3D}=1.87$ from our full self-consistent
RPA calculations (thick solid line) and from the hydrodynamic Eq.~(\ref{lh})
(thin solid line). Far inside the solid, our full calculation is close to the
hydrodynamic prediction (see also Fig.~\ref{fig2}) and coincides with the
result one obtains from the bulk screened interaction of Eq.~(\ref{bulk})
(horizontal dashed line). Near
the surface, our full calculation considerably deviates from the hydrodynamic
prediction and converges far outside the solid with the asymptotic curve
dictated by Eq.~(\ref{wout2}) with $d_\perp(0)=0.2\,\lambda_F$ (dotted
line).\cite{notenew}

At this point, it is interesting to note that within a local picture of the 3D
response the long-wavelength $I(z_d)$ screened interaction would be zero for
all locations of the 2D sheet inside the metal ($z_d\leq 0$), showing that the
screening of 2D electron-density oscillations would be complete and no
acoustic surface plasmon would occur. It is precisely the nonlocality of the
3D response (finite values of the 3D momentum $k$ are still present in the 2D
long-wavelength limit) which provides incomplete screening and allows,
therefore, the formation of acoustic surface plasmons in the interior of the
solid. We also note that within a simple nonlocal picture of the 3D response,
such as the hydrodynamic and specular-reflection models described above, the
screening of 2D electron-density oscillations would still be complete at the
jellium edge ($z_d=0$). Hence, in the real situation where the 2D
surface-state band is located very near the jellium edge the occurence of
acoustic surface plasmons is originated by a combination of the nonlocality of
the 3D response and the spill out of the 3D electron density into the vacuum.

\begin{figure}
\includegraphics[width=0.45\textwidth,height=0.3375\textwidth]{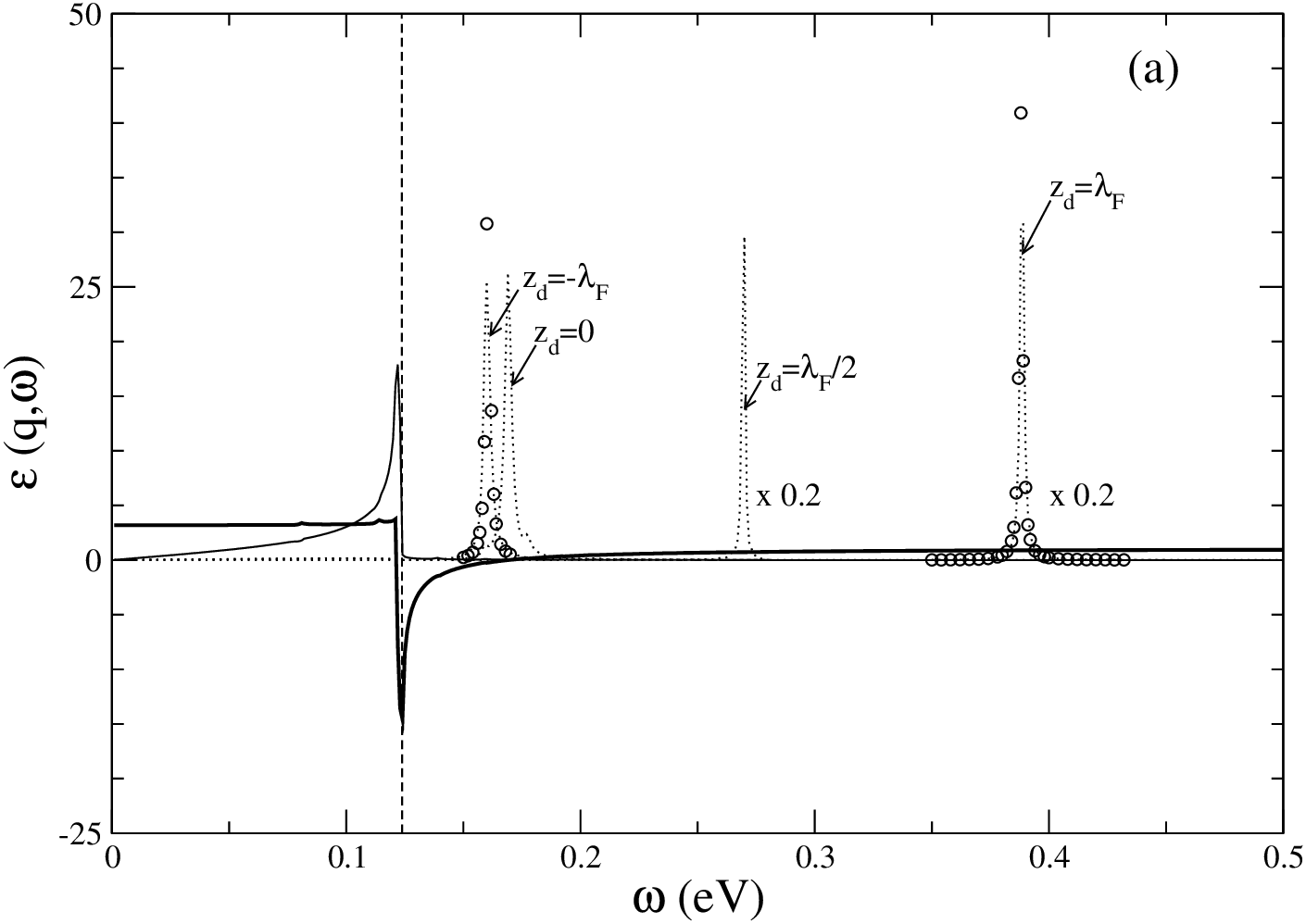}\\
\vspace*{1.2cm}
\includegraphics[width=0.45\textwidth,height=0.3375\textwidth]{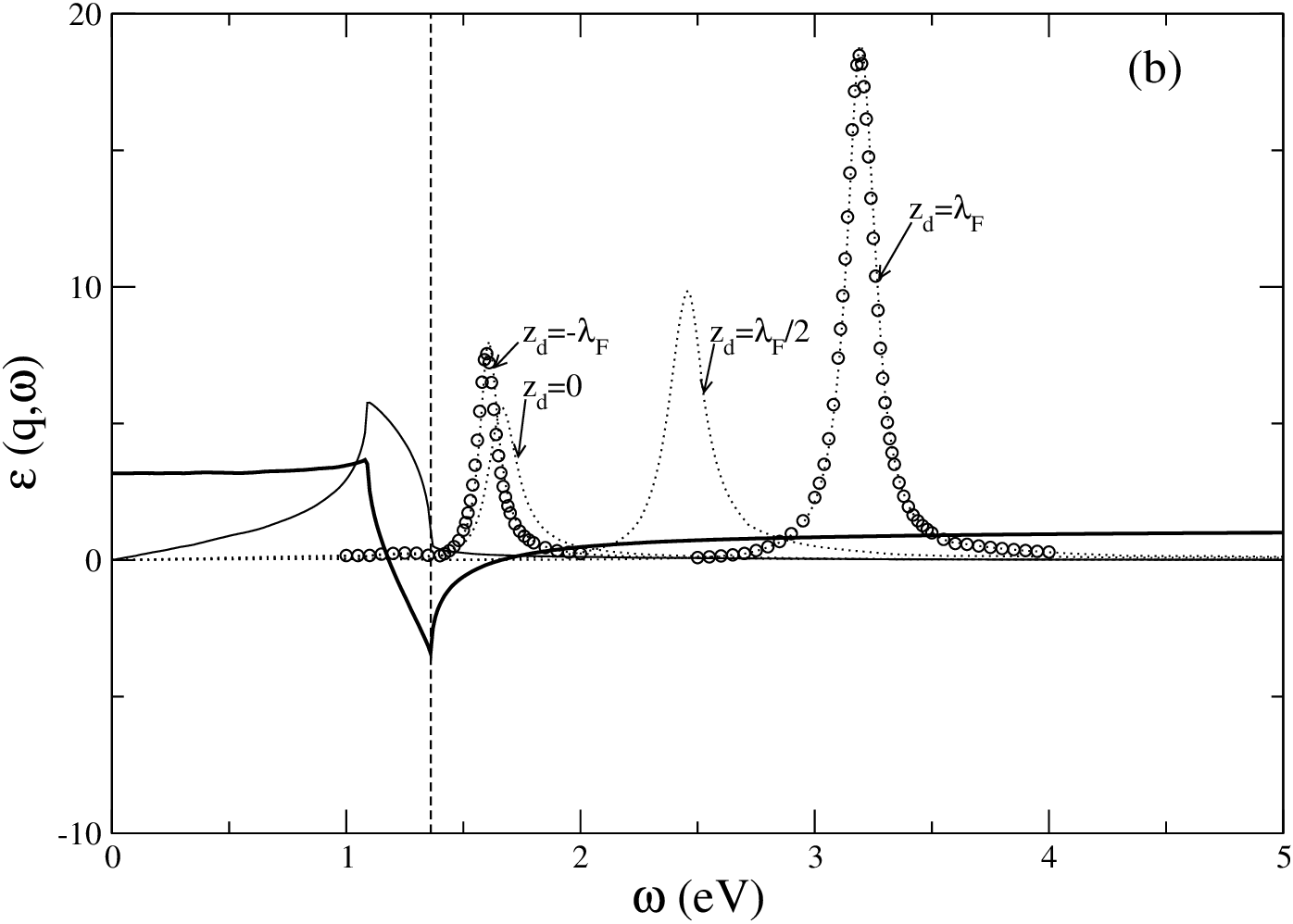}\\
\caption{As in Fig.~\ref{fig1}, but now for a 2D sheet that is located at the
jellium edge ($z_d=0$). Also shown is the effective 2D energy-loss function
${\rm Im}\left[-\epsilon_{eff}^{-1}(q,\omega)\right]$ for $z_d=-\lambda_F$,
$z_d=\lambda_F/2$, and $z_d=\lambda_F$ (dotted lines). The open circles
represent the effective 2D energy-loss function ${\rm
Im}\left[-\epsilon_{eff}^{-1}(q,\omega)\right]$ obtained from the limiting
Eq.~(\ref{bulk}) appropriate for a 2D sheet far inside the metal and from the
limiting Eq.~(\ref{g}) with $z_d=\lambda_F$ appropriate for a 2D sheet far
outside the metal. These calculations are found to coincide with the full
calculations for $z_d=-\lambda_F$ and $z_d=\lambda_F$,
respectively. As in Fig.~\ref{fig3}(a), the calculations presented here for
$z_d=\lambda_F$ and $q=0.01\,a_0^{-1}$ have been carried out by replacing the
energy $\omega$ by a complex quantity $\omega+i\eta$ with $\eta=0.05\,{\rm
eV}$. All remaining calculations have been carried out for real frequencies,
i.e., with $\eta=0$.\label{fig4}}
\end{figure}

Figs.~\ref{fig4}a and \ref{fig4}b exhibit the results we have obtained for the
effective dielectric function of Eq.~(\ref{eff2}) [with $q=0.01\,a_0^{-1}$
(Fig.~\ref{fig4}a) and $q=0.1\,a_0^{-1}$ (Fig.~\ref{fig4}b)]  by using the full
2D noninteracting density-response function, $\chi_{2D}^0(q,\omega)$, and the
self-consistent
RPA screened interaction, $W(z_d,z_d;q,\omega)$, with electron-density
parameters $r_s^{2D}=3.14$ and $r_s^{3D}=1.87$ corresponding to Be(0001). In
these figures the 2D sheet has been taken to be located at $z_d=0$, as
approximately occurs with the quasi-2D surface-state band in Be(0001). For
comparison, also shown in these figures are the results we have obtained for
the energy-loss function when the 2D sheet is located inside the metal
at $z_d=-\lambda_F$ and outside the metal at $z_d=\lambda_F/2$ and
$z_d=\lambda_F$.

An inspection of Fig.~\ref{fig4} shows that (i) the results we have obtained
for
$z_d=-\lambda_F$ and $z_d=\lambda_F$ are exactly reproduced by the limiting
Eqs.~(\ref{bulk}) and (\ref{g}) appropriate for a 2D sheet far inside and far
outside the metal surface, respectively; and (ii) in the actual situation
where the 2D surface-state band is located very near the jellium positive
background edge ($z_d=0$), a well-defined low-energy acoustic plasmon occurs,
the sound velocity being very close to the limiting case of a 2D sheet far
inside the metal surface and being, therefore, just above $\omega_u$. This is
in agreement with the recent prediction that in a real metal
surface where a partially occupied quasi-2D surface-state band {\it coexists}
in the same region of space with the underlying 3D continuum an acoustic
surface plasmon should occur at energies just above the upper edge of the 2D
e-h pair continuum.\cite{silkin}

\begin{figure}
\includegraphics[width=0.45\textwidth,height=0.3375\textwidth]{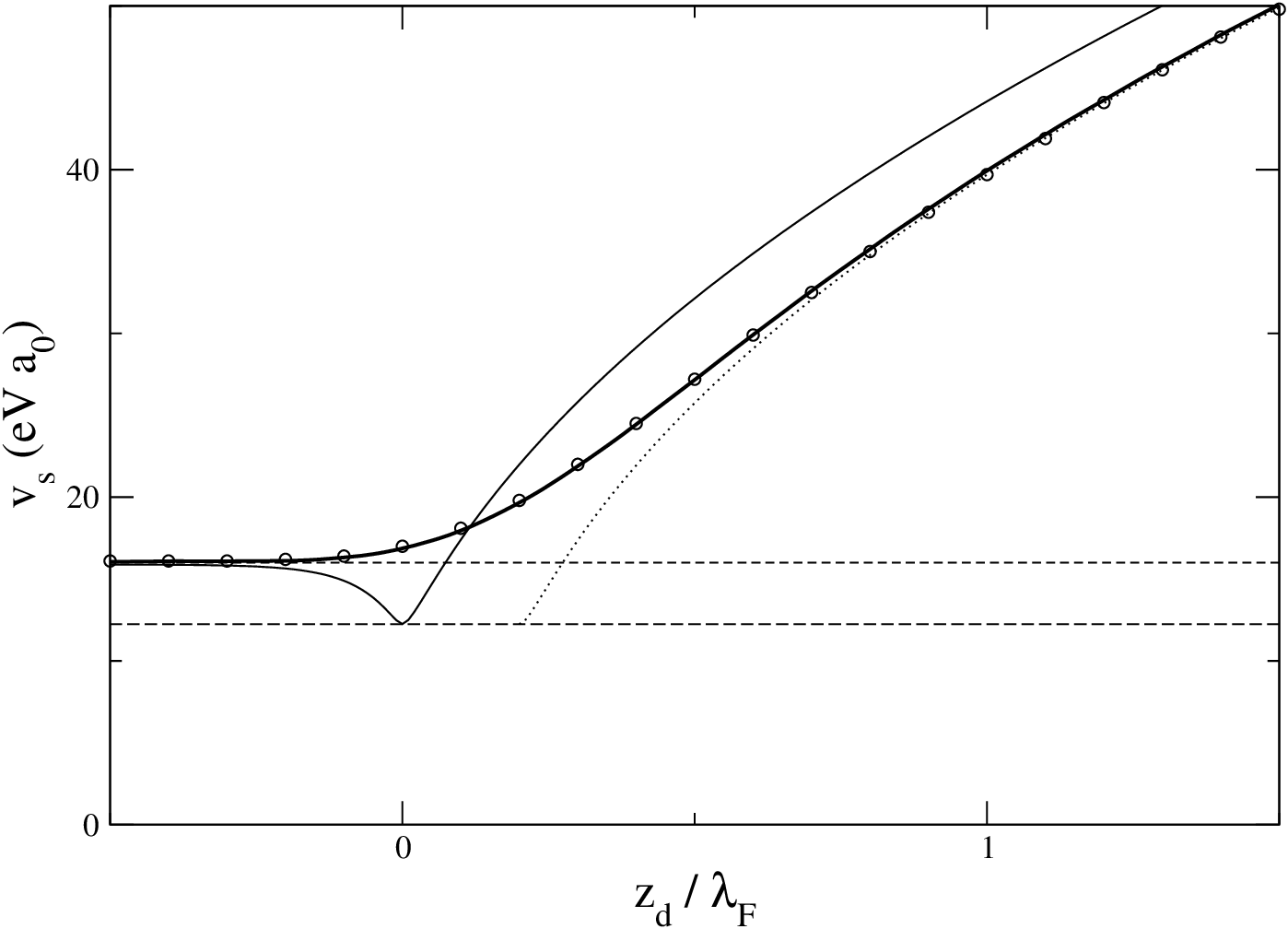}
\caption{The open circles represent the sound velocity $v_s$ ($\omega=v_s\,q$)
of the low-energy acoustic plasmon that is visible in Fig.~\ref{fig4} versus
the location $z_d$ of the 2D sheet
with respect to the jellium edge. The horizontal short-dashed line represents
the
result we have obtained from the limiting Eq.~(\ref{bulk}) appropriate for a
2D sheet far inside the metal. The dotted line represents the result we have
obtained from the limiting Eq.~(\ref{alpha4}) with $d_\perp=0.2\,\lambda_F$,
which is appropriate for a 2D sheet far outside the metal. The
long-wavelength limit $v_F$ of the upper edge $\omega_u/q$ of the 2D e-h pair
continuum is represented by an horizonal long-dashed line. The
thick and thin solid lines represent the results obtained from
Eq.~(\ref{alpha2}) by using the actual RPA $I(z_d)$ and the hydrodynamic
Eq.~(\ref{lh}), respectively. 2D and 3D
electron densities have been taken to be those corresponding to the Wigner
radii $r_s^{2D}=3.14$ and $r_s^{3D}=1.87$, respectively. The 2D effective mass
has been taken to be $m=1$.
\label{fig5}}
\end{figure}

The sound velocity $v_s$ ($\omega=v_s\,q$) of the acoustic plasmon that is
visible in Fig.~\ref{fig4} is displayed in Fig.~\ref{fig5} versus the location
$z_d$ of the 2D sheet relative to the jellium edge, as obtained from our full
RPA self-consistent calculation of the effective 2D dielectric function of
Eq.~(\ref{eff2}) (open circles), together with the sound velocity
$v_s=\alpha\,v_F$ obtained from Eq.~(\ref{alpha4}) with
$d_\perp(0)=0.2\,\lambda_F$ (dotted line). When the 2D sheet is located inside
the metal surface, the
sound velocity nicely converges with the RPA bulk calculation from
Eq.~(\ref{bulk}) (horizontal short-dashed line). When the 2D
sheet is located outside the metal surface, the sound velocity converges with
the limiting value $\alpha\,v_F$ obtained from Eq.~(\ref{alpha4}) and 
$d_\perp(0)=0.2\,\lambda_F$. For comparison, also shown in this figure is the
result we have obtained from Eq.~(\ref{alpha2}) by using the actual RPA
$I(z_d)$ screened interaction (thick solid line) and from the hydrodynamic
Eq.~(\ref{lh}). These calculations clearly show that Eq.~(\ref{alpha2})
accurately reproduces the dispersion of acoustic surface plasmons, as long as
the long-wavelength limit $I(z_d)$ of the screened interaction is described
self-consistently with full inclusion of the electronic selvage structure at
the surface.  

\begin{figure}
\vspace*{1.2cm}
\includegraphics[width=0.45\textwidth,height=0.3375\textwidth]{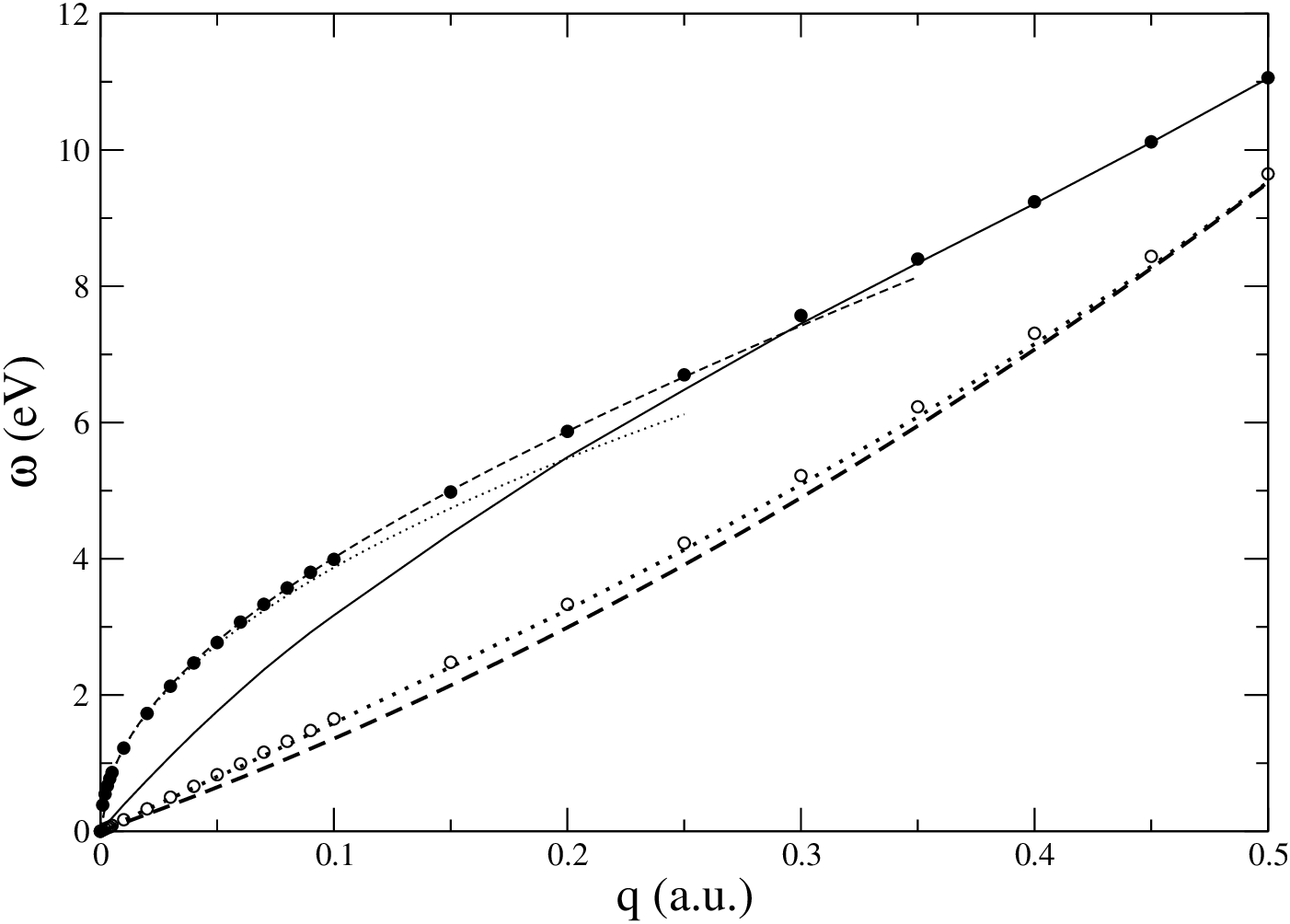}
\caption{Dispersion of the acoustic plasmon occurring in a 2D sheet that is
taken to be
located far inside the solid (thick dotted line), at $z_d=0$ (open circles),
at $z_d=\lambda_F$ (solid line), and infinitely far outside the metal (solid
circles). The thick dashed line represents the upper edge
$\omega_u=v_Fq+q^2/(2m)$ of the 2D e-h pair continuum. The thin dotted line
represents the 2D plasmon energy $\omega_{2D}$ dictated by Eq.~(\ref{2d}),
which is accurate at long wavelengths ($q\to 0$). The thin dashed line
represents the 2D plasmon energy $\sqrt{\omega_{2D}^2+3v_F^2q^2/4}$ that is
obtained after an expansion of $\chi_{2D}^0(q,\omega)$ in powers of
$v_Fq/\omega$. 2D and 3D electron densities have been taken
to be those corresponding to the Wigner radii $r_s^{2D}=3.14$ and
$r_s^{3D}=1.87$, respectively. The 2D effective mass has been taken to be
$m=1$.
\label{fig6}}
\end{figure}

The sound velocity of Fig.~\ref{fig5} (open circles) has been obtained from the
effective 2D
dielectric function at very low 2D momenta, where the energy of
the acoustic plasmon is linear in $q$. The behaviour of this plasmon energy as
a function of the 2D momentum $q$ is displayed in Fig.~\ref{fig6}, with the 2D
sheet chosen to be located far inside the solid (thick dotted line), at $z_d=0$
(open circles), at $z_d=\lambda_F$ (solid line), and infinitely far outside
the solid
(solid circles). The upper edge of the 2D e-h pair continuum is represented by
a thick dashed line, showing that in the real situation where the 2D sheet is
located near the jellium edge the energy of the acoustic surface plasmon (open
circles) is just outside the 2D e-h pair continuum for all momenta under
study.

\section{Summary and Conclusions}

The partially occupied band of Shockley surface states in a variety of metal
surfaces is known to form a quasi-2D electron gas that is immersed in a
semiinfinite 3D gas of valence electrons. In order to describe the impact of
the dynamical screening of the semi-infinite 3D continuum on the electronic
excitations at the 2D electron gas of Shockley surface states, we have
presented a model in which the dynamical screening of 3D valence electrons is
incorporated through the introduction of an effective 2D dielectric function.

We have considered the two limiting cases in which the 2D sheet is
located far inside and far outside the metal surface. In both cases, the
dynamical screening of the valence electrons in the metal is found to change
the 2D plasmon energy from its characteristic square-root behaviour to a
linear dispersion, the sound velocity being proportional to the Fermi momentum
of the 2D gas. As this collective oscillation occurs in a region of 2D
momentum space where 2D e-h pairs cannot be produced, this is a well-defined
acoustic plasmon. The finite width of the plasmon peak is due to a
small probability for the plasmon to decay into e-h pairs of the 3D substrate.

We have shown explicitly that when the 2D sheet {\it coexists} in the same
region of space with the underlying 3D continuum the origin of acoustic
surface plasmons, which have been overlooked over the years, is dictated by a
combination of the nonlocality of the 3D response and the spill out of  the 3D
electron density into the vacuum, both providing incomplete screening of the
2D electron-density oscillations.

We have carried out self-consistent DFT calculations of the dynamical
density-response function of the 3D system of valence electrons, and we have
found that a well-defined acoustic plasmon exists for all possible locations
of the 2D sheet relative to the metal surface. The energy dispersion of this
acoustic surface plasmon is slightly higher than the energy of the collective
excitation that has recently been predicted to exist at real metal surfaces
where a quasi-2D surface-state band coexists with the underlying 3D
continuum.\cite{silkin} Small differences between the plasmon energies
obtained here and those reported previously\cite{silkin} are due to the
absence in the present model of transitions between 2D and 3D
states.\cite{silkin2}

\acknowledgments

Partial support by the University of the Basque Country, the Basque
Unibertsitate eta Ikerketa Saila, the Spanish MCyT, and the Max Planck Research
Award Funds is gratefully acknowledged. V.U.N. acknowledges support by the
Korea Research Foundation through Grant No. KRF-2003-015-C00214 and the
hospitality of the Donostia International Physics Center (DIPC).

\appendix

\section{Specular-reflection model of the 3D response}\label{ap2}

Either by assuming that electrons are specularly reflected at the surface
(SRM)\cite{ritchie2} or by invoking the so-called classical infinite-barrier
model (CIBM) of a jellium surface,\cite{griffin,nazarov} one finds
\begin{widetext}
\begin{equation}\label{srm}
W(z_d,z_d;q,\omega)=v_q\,\cases{1-{\rm
e}^{-2qz_d}\left[1-\epsilon_s(0;\omega)\right]/
\left[1+\epsilon_s(0;\omega)\right],&$z_d\geq 0$\cr\cr
\epsilon_s(0;q,\omega)+\epsilon_s(2z_d;q,\omega)-2
\epsilon_s^2(z_d;q,\omega)/\left[\epsilon_s(0;q,\omega)+1\right],&elsewhere},
\end{equation}
\end{widetext}
where
$k=\sqrt{q^2+q_z^2}$ is a 3D momentum, and
\begin{equation}\label{srm2}
\epsilon_s(z;q,\omega)={q\over\pi}\int_{-\infty}^{+\infty}{dq_z\over
k^2}\,{\rm e}^{iq_zz}\,\epsilon_{3D}^{-1}(k,\omega),
\end{equation}
$\epsilon_{3D}(q,\omega)$ being the dielectric function of a uniform (and
infinite) 3D electron gas.

If the 3D dielectric function  $\epsilon_{3D}(q,\omega)$ is chosen to be the
hydrodynamic dielectric function of Eq.~(\ref{hydro}), then one finds
\begin{equation}
\epsilon_s(z;q,\omega)={1\over
\omega^2-\omega_p^2}\left[\omega^2-{\beta\omega_p^2 q\over
\sqrt{\beta^2q^2+\omega_p^2-\omega^2}}\right]\,e^{-q|z_d|},
\end{equation}
which in combination with Eq.~(\ref{srm}) yields Eq.~(\ref{lh}).


\begin{thebibliography}{100}
\bibitem{note00} The sound velocity of this acoustic mode is, however, close to
the Fermi velocity of the 2D surface-state band, which is typically a few
orders of magnitude larger than the sound velocity of acoustic phonons in
metals.
\bibitem{note01} If one replaces the dielectric function
$\epsilon_{3D}(\omega)$ entering Eq.~(\ref{local}) by a constant $\kappa$ and
expands $\chi_{2D}^0(q,\omega)$ in powers of $v_Fq/\omega$, Eq.~(\ref{zero})
is easily found to yield the plasmon dispersion
$\omega^2=\omega_{2D}^2/\kappa+3v_F^2q^2/4$ (see Ref.~\onlinecite{stern}). This
suggests that at low frequencies, where the dielectric constant $\kappa$ is
large, an acoustic plasmon should be expected to occur at
$\omega\approx\sqrt{3/4}\,v_Fq$; however, at these low frequencies, where
$\omega<v_Fq$, the plasmon dispersion
$\omega^2=\omega_{2D}^2/\kappa+3v_F^2q^2/4$ fails. Furthermore, a careful
analysis of the long-wavelength behaviour of $\chi_{2D}^0(q,\omega)$ shows
that an infinitely large dielectric constant $\kappa$ yields an effective 2D
dielectric function which has no zeros, so that no acoustic plasmon occurs in
this model.
\bibitem{note0} The value $\beta=\sqrt{3/5}\,k_F$ is usually chosen to describe
processes in which high frequencies of the order of the 3D plasma frequency
are involved. For the low energies of interest here, however, the hydrodynamic
value $\beta=\sqrt{1/3}\,k_F$ should be more appropriate.
\bibitem{note1} On Be(0001), an occupied surface state with
a binding energy of $2.8\,{\rm eV}$ at $\bar\Gamma$ exists in the
$\Gamma_3^+-\Gamma_4^-$ bulk band gap. This surface state disperses with
momentum parallel to the surface, thereby forming a surface-state band with a
2D Fermi energy $\varepsilon_F=2.8\,{\rm eV}$. If one takes, for simplicity,
the effective mass of surface-state electrons to be equal to the free-electron
mass, the 2D Fermi momentum $q_F$ and Wigner radius $r_s^{2D}$
[$r_s^{2D}=\sqrt{2}/(q_Fa_0)$] are found to be $q_F=0.45a_0^{-1}$ and
$r_s^{2D}=3.14$, respectively. On the other hand, Be has two valence electrons
per atom, which yields an average 3D electron density $\bar n$ corresponding
to the 3D Wigner radius $r_s^{3D}=1.87$.
\bibitem{loss} The energy-loss function ${\rm
Im}\left[-\epsilon_{eff}^{-1}(q,\omega)\right]$ is not necessarily positive
definite and may not give the true absorptive response of the combined 2D and
3D systems; however, its resonant structure is a true reflection of the modes
of the system.
\bibitem{z0} $z_0$ is chosen sufficiently large for the physical results to be
insensitive to the precise value employed.
\bibitem{notenew} The image-plane coordinate $d_\perp(0)=0.2\lambda_F$ has been chosen in such a way that Eq.~(42) reproduces the asymptotic behaviour of the full RPA calculation represented in Fig.~4 by a thick solid line.
\end{thebibliography}
\end{document}